\documentclass[11pt]{article}

\usepackage{amssymb}
\usepackage{amsmath}
\usepackage{graphicx}
\usepackage{color}
\usepackage{multirow}
\usepackage{fullpage}

\usepackage{slashbox}

\begin{document}

\newtheorem{theorem}{Theorem}[section]
\newtheorem{lemma}{Lemma}[section]
\newtheorem{corollary}{Corollary}[section]
\newtheorem{claim}{Claim}[section]
\newtheorem{proposition}{Proposition}[section]
\newtheorem{definition}{Definition}[section]
\newtheorem{fact}{Fact}[section]
\newtheorem{example}{Example}[section]

\newcommand{\quod}{\hfill $\blacksquare$ \bigbreak}
\newcommand{\reals}{I\!\!R}
\newcommand{\property}{I\!\!P}
\newcommand{\np}{\mbox{{\sc NP}}}
\newcommand{\sing}{\mbox{{\sc Sing}}}
\newcommand{\prob}{\mbox{Prob}}
\newcommand{\atm}{\mbox{{\sc ATM}}}
\newcommand{\hopn}{\hop_{\cN}}
\newcommand{\atmn}{\atm_{\cN}}
\newcommand{\cA}{{\cal A}}
\newcommand{\cO}{{\cal O}}
\newcommand{\cR}{{\cal R}}
\newcommand{\cP}{{\cal P}}
\newcommand{\cC}{{\cal C}}
\newcommand{\C}{{\cal C}}
\newcommand{\cB}{{\cal B}}
\newcommand{\cG}{{\cal G}}
\newcommand{\cN}{{\cal N}}
\newcommand{\cU}{{\cal U}}
\newcommand{\cF}{{\cal F}}
\newcommand{\cT}{{\cal T}}
\newcommand{\hx}{\hat{x}}
\newcommand{\cS}{{\cal S}}

\baselineskip    0.2in
\parskip         0.1in
\parindent       0.0in

\newcommand{\RVN}{\mathop{RVC}}
\newcommand{\RV}{\mathop{RV}}
\newcommand{\RVO}{\mathop{RVO}}
\newcommand{\RVMN}{\mathop{RVCM}}
\newcommand{\RVMO}{\mathop{RVMO}}
\newcommand{\RVM}{\mathop{RVM}}

\newcommand{\qed}{\hfill $\square$ \smallbreak}
\newenvironment{proof}{\noindent{\bf Proof:}}{\qed}
\newenvironment{proofof}{\noindent{\bf Proof of}}{\qed}

\bibliographystyle{plain}

\title{{\bf Asynchronous deterministic rendezvous in bounded terrains}}

\author{
Jurek Czyzowicz  \footnotemark[1]  \footnotemark[2]
\and
David Ilcinkas \footnotemark[3]
\and
Arnaud Labourel \footnotemark[1] \footnotemark[4]
\and
Andrzej Pelc \footnotemark[1] \footnotemark[5]
}
\date{ }
\maketitle
\def\thefootnote{\fnsymbol{footnote}}

\footnotetext[1]{ \noindent
D\'epartement d'informatique, Universit\'e du Qu\'ebec en Outaouais, Gatineau,
Qu\'ebec J8X 3X7, Canada.
E-mails: {\tt jurek@uqo.ca}, {\tt labourel.arnaud@gmail.com}, {\tt pelc@uqo.ca}  
}
\footnotetext[2]{ \noindent
Partially supported by NSERC discovery grant.
}
\footnotetext[3]{ \noindent
LaBRI, Universit\'{e} Bordeaux I,
33405 Talence, France.
E-mail: {\tt david.ilcinkas@labri.fr}
}
\footnotetext[4]{ \noindent
This work was done during this author's stay at the
Universit\'e du Qu\'{e}bec en Outaouais as a postdoctoral fellow.  
} 
\footnotetext[5]{ \noindent  
Partially supported by NSERC discovery grant and     
by the Research Chair in Distributed Computing at the
Universit\'e du Qu\'{e}bec en Outaouais. 
}

\begin{abstract}
Two mobile agents (robots) have to meet in an a priori
unknown bounded terrain modeled as a polygon, possibly with polygonal obstacles.
Agents are modeled as points, and each 
of them is equipped with a compass. Compasses of agents may be incoherent.
Agents construct their routes, but
the actual walk of each agent is decided by the adversary: the movement of the agent can be at arbitrary speed, the agent may sometimes stop or go back and forth, as long as the walk of the agent in each segment of its route is continuous, does not leave it and covers all of it. 
We consider several scenarios, depending on three factors:
(1) obstacles in the terrain are present, or not, (2) compasses of both agents agree, or not, (3) agents have or
do not have a map of the terrain with their positions marked. The cost of a rendezvous algorithm
is the worst-case sum of lengths of the agents' trajectories until their meeting.
For each scenario we design a deterministic rendezvous algorithm and analyze its cost. We also
prove lower bounds on the cost of any deterministic rendezvous algorithm in each case. For all scenarios
these bounds are tight.

\vspace*{1cm}

\noindent
{\bf keywords:} mobile agent, rendezvous, deterministic, polygon, obstacle 
\vspace*{5cm}

\end{abstract}

\thispagestyle{empty}

\pagebreak

\section{Introduction}

{\bf The problem and the model.}
Two mobile agents (robots) modeled as points starting at different locations of an a priori unknown bounded terrain have to meet. The terrain is represented as a polygon possibly with a finite number of polygonal obstacles. We assume that the boundary of the terrain is included in it. Thus, formally, a terrain is a set $\cP_0 \setminus (\cP_1 \cup \dots \cup \cP_k)$, where $\cP_0$ is a closed polygon and $\cP_1,\dots , \cP_k$ are disjoint open polygons included in $\cP_0$. We assume that an agent knows if it is at an interior or at a boundary point, and in the latter case it is capable of walking along the boundary in both directions (i.e., it knows the slope(s) of the boundary at this point).  However, an agent cannot sense the terrain or the other agent at any vicinity of its current location. Meeting (rendezvous) is defined as the equality of points representing agents at some moment of time.

We assume that each agent has a unit of length and a compass. Compasses of agents may be incoherent, however we assume that agents have the same (clockwise) orientation of their system of coordinates. An additional tool, which may or may not be available to the agents, is a map of the terrain. The map available to an agent is scaled (i.e., it accurately shows the distances), distinguishes the starting positions of this agent and the other one, and is oriented according to the compass of the agent. (Hence maps of different agents may have different North.)

All our considerations concern deterministic algorithms.
The crucial notion is the {\em route} of the agent which is a finite polygonal path in the terrain.
The adversary initially places an agent at some point in the terrain. The agent constructs its route in steps in the following way. In every step the agent starts at some point $v$; in the first step, $v$ is the starting point chosen by the adversary. The agent chooses a direction $\alpha$, according to its compass, and a distance $x$. If the segment of length $x$ in direction $\alpha$ starting in $v$ does not intersect the boundary of the terrain, the step ends when the agent reaches point $u$ at distance $x$ from $v$ in direction $\alpha$. Otherwise, the step ends at the closest point of the boundary in direction $\alpha$. If the starting point $v$ in a step
is in a segment of the boundary of the terrain, the agent has also an option (in this step) to follow this segment of the boundary in any of the two directions till its end or for some given distance along it. Steps are repeated until rendezvous,
or until the route of the agent is completed. 

We consider the  {\em asynchronous} version of the rendezvous problem.
The asynchrony of the agents' movements is captured by the assumption 
that the actual walk of each agent is decided by the adversary: the movement of the agent can be at arbitrary speed, the agent may sometimes stop or go back and forth, as long as the walk of the agent in each segment of its route is continuous, does not leave it and covers all of it. More formally, the route in a terrain is a sequence $(S_1,S_2,\dots, S_k)$ of segments, where $S_i=[a_i, a_{i+1}]$ is the segment corresponding to step $i$.
In our algorithms the route is always finite. This means that the agent stops at some point, regardless of the moves of the other agent. We now describe the walk $f$ of an agent on its route. Let $R=(S_1,S_2,\dots,S_k)$ be the route of an agent. Let $(t_1,t_2,\dots,t_{k+1})$, where $t_1=0$, be an increasing sequence of reals, chosen by the adversary, that represent points in time. Let $f_i:[t_i,t_{i+1}]\rightarrow [a_i,a_{i+1}]$ be any continuous function, chosen by the adversary, such that $f_i(t_i)=a_i$ and $f_i(t_{i+1})=a_{i+1}$. For any $t\in [t_i,t_{i+1}]$, we define $f(t)=f_i(t)$. 
The interpretation of the walk $f$ is as follows: at time $t$ the agent is at the point $f(t)$ of its route and after time $t_{k+1}$ the agent remains inert. This general definition of the walk and the fact that it is constructed by the adversary capture the asynchronous characteristics of the process.  Throughout the paper, {\em rendezvous} means deterministic asynchronous  rendezvous.

Agents with routes $R_1$ and $R_2$ and with walks $f_1$ and $f_2$ meet at time $t$,
if points $f_1(t )$ and $f_2(t )$ are equal. A rendezvous is guaranteed for routes $R_1$ and $R_2$, if the agents using these routes meet at some time $t$,  regardless of the walks chosen by the adversary. The trajectory of an agent is the sequence of segments on its route until rendezvous. (The last segment of the trajectory of an agent 
may be either the last segment of its route or any of its segments or a portion of it, if the other agent 
is met there.) The cost of a rendezvous algorithm is the worst case sum of lengths of segments of trajectories of both agents, where the worst case is taken over all terrains with the considered values of parameters, and all adversarial decisions.

We consider several scenarios, depending on three factors:
(1) obstacles in the terrain are present, or not, (2) compasses of both agents agree, or not, (3) agents have or do not have a map of the terrain. Combinations of the presence or absence of these factors give rise to eight scenarios. For each scenario we design a deterministic rendezvous algorithm and analyze its cost. We also prove lower bounds on the cost of any deterministic rendezvous algorithm in each case. For all scenarios these bounds are tight. 

One final precision has to be made. For all scenarios except those with incoherent compasses and the presence of obstacles (regardless of the availability of a map), 
agents may be anonymous, i.e., they execute identical algorithms. By contrast, with the presence of obstacles and incoherent compasses, anonymity would preclude feasibility of 
rendezvous in some situations. Consider a square with one square obstacle positioned at its center. Consider two agents starting at opposite (diagonal) corners
of the larger square, with compasses pointing to opposite North directions. If they execute identical algorithms and walk at the same speed, then at each time they are in symmetric positions in the terrain and hence rendezvous is impossible.
The only way to break symmetry for a deterministic rendezvous in this case is to equip the agents with distinct labels (which are positive integers). Hence, this is the assumption we make for the scenarios with the presence of obstacles and incoherent compasses
(both with and without a map). 
For any label $\mu$, we denote by $|\mu|$ the length of the binary representation of the label, i.e., $|\mu|=\lfloor\log\mu\rfloor+1$.

{\bf Our results.}
The cost of our algorithms depends on some of the following parameters (different parameters for different scenarios, see the discussion in Section 4):
$D$ is the distance between starting positions of agents in the terrain (i.e., the length of the shortest path between them included in the terrain), $P$ is the perimeter of the terrain, 
(i.e., the sum of perimeters of all polygons $\cP_0,\cP_1,\dots ,\cP_k$), $x$ is the largest perimeter of an obstacle, and $l$ and $L$ are the smaller and larger labels of agents, respectively, for the two scenarios that require different labels, as remarked above., i.e., for the scenarios
with the presence of obstacles and incoherent compasses.

Our rendezvous algorithms rely on two different ideas: either meeting in a uniquely defined point of the terrain, or meeting on a uniquely defined cycle. It turns out that a uniquely defined point can be found in all scenarios except those with the presence of obstacles and incoherent compasses.  In this case even anonymous agents can meet. On the other hand, with the presence of obstacles and incoherent compasses, 
such a uniquely defined point may not exist, as witnessed by the above quoted example of a square with one square obstacle positioned at its center. For these scenarios we resort to the technique of meeting at a common cycle, breaking symmetry by different labels of agents.

We first summarize our results concerning rendezvous when each of the agents is equipped with a map showing its own position and that of the other agent. If compasses of the agents are coherent, then we show a rendezvous algorithm at cost $D$, which is clearly optimal. Otherwise, and if the terrain does not contain obstacles, then we show an algorithm whose cost is again $D$, and hence optimal. Finally, with incoherent compasses in the presence of obstacles,
we show a rendezvous algorithm at cost $O(D|l|)$; in the latter scenario we show that cost $\Omega(D|l|)$ is necessary for some terrains.

Our results concerning rendezvous without a map are as follows.
If compasses of the agents are coherent, then we show a rendezvous algorithm
at cost $O(P)$. We also show a matching lower bound $\Omega(P)$ in this case.
If compasses of the agents are incoherent, but the terrain does not contain obstacles, then we show
a rendezvous algorithm at cost $O(P)$ and again a matching lower bound $\Omega(P)$.
Finally, in the hardest of all scenarios (presence of obstacles, incoherent compasses
and no map) we have a rendezvous algorithm at cost $O(P+x|L|)$ and a matching lower bound
$\Omega (P+x|L|)$. Table~\ref{tab:results} summarizes our results. 

\begin{table}[ht]
{\small
\begin{tabular}{|c|c|c||c|c|c|}\hline
\multicolumn{3}{|c||}{Rendezvous with a map} & \multicolumn{3}{c|}{Rendezvous without a map}\\ \hline
\backslashbox{obstacles}{compasses}& coherent & incoherent &\backslashbox{obstacles}{compasses} & coherent & incoherent\\
\hline
no & \multirow{2}{*}{$D$} & $D$ & no & \multirow{2}{*}{$\Theta(P)$} & $\Theta(P)$\\
\cline{1-1}\cline{3-4}\cline{6-6}
yes &  & $\Theta(D| l |)$ & yes & & $\Theta(P+x| L|)$\\
 \hline
\end{tabular}
}
\caption{Summary of results}
\label{tab:results}
\end{table}


{\bf Related work.}
The rendezvous problem was first described in \cite{schelling60}. A detailed discussion of the large 
literature on rendezvous can be found in the excellent book
\cite{alpern02b}. Most of the results in this domain can be divided
into two classes: those 
considering the geometric scenario (rendezvous in the line, see, e.g., 
\cite{baston98,baston01,gal99,Sta09},
or in the plane, see, e.g., \cite{anderson98a,anderson98b}), and those
discussing rendezvous in graphs,
e.g., \cite{alpern95a, alpern99}. Some of the authors, e.g.,
\cite{alpern95a,alpern02a,anderson90,baston98,israeli} consider
the probabilistic scenario where inputs and/or rendezvous strategies are random. 
Randomized rendezvous strategies use random walks in
graphs, which
were thoroughly investigated and applied also to other problems, such as graph traversing
\cite{akllr}, on-line algorithms
\cite{cdrs} and estimating volumes of convex bodies \cite{dfk}.
A generalization of the rendezvous
problem is that of gathering \cite{fpsw,israeli,KKN,KMP,lim96,thomas92}, when more than
two agents have to meet in one location.

If graphs are unlabeled, deterministic rendezvous requires breaking symmetry, which can be accomplished either
by allowing marking nodes or by labeling the agents.
Deterministic rendezvous with anonymous agents working in unlabeled graphs but 
equipped with tokens used to mark nodes was considered e.g., in~\cite{KKSS}.
In~\cite{YY} the authors studied the task of gathering many agents
with unique labels. In \cite{DFKP,KM,TZ} deterministic rendezvous in graphs with
labeled agents was considered.
However, in all the above papers, the synchronous setting was assumed.
Asynchronous gathering under geometric
scenarios has been studied, e.g., in \cite{CFPS,fpsw,Pr} in different models than ours:
agents could not remember past events, but  
they were assumed to have at least partial visibility of the scene.
The first paper
to consider deterministic asynchronous rendezvous in graphs was \cite{DGKKPV}.
The authors concentrated on complexity of rendezvous in simple graphs, such as the ring
and the infinite line. They also showed feasibility of  deterministic asynchronous rendezvous in arbitrary finite
connected graphs with {\em known} upper bound on the size. Further improvements of the above results 
for the infinite line were proposed in~\cite{Sta09}. Gathering many robots in a graph, under a  different 
asynchronous model and assuming that the whole graph is seen by each robot, has been studied in \cite{KKN,KMP}.

\section{Rendezvous with a map}

We start by describing the following procedure that finds a unique shortest path from the starting position of one agent to the other.
The procedure works in all scenarios in which agents have a map of the terrain with their positions indicated.

\fbox{
   \begin{minipage}{0.98\textwidth}
\begin{tabbing}
\textbf{Pro}\=\textbf{cedure} path {\tt UniquePath}(point $v$, point $w$)\\
1 \> point $u:=v$; path $p:=\{v\}$;\\
2 \>$\cS=\{p_s\mid p_s \mbox{ is a shortest path between $v$ and $w$}\}$;\\
3 \> \textbf{whi}\=\textbf{le} $(u\neq w)$ \textbf{do}\\
4 \>\> $\cU:=$\= all paths $p_s$ of $\cS$ such that the first segment of 
the subpath of $p_s$ leading from $u$\\ 
\>\>\> to $w$ is the first in clockwise order around $u$ starting from the direction $vw$;\\
5 \>\> $p':=\bigcap_{p_s\in \cU}p_s;$\\
6 \>\> extend $p$ with the connected part of $p'$ containing $u$;\\
7 \> \> $u:=$ new end of path $p$;\\
8\> \textbf{return} $p$;
\end{tabbing}
\end{minipage}
}

\begin{lemma}\label{lem:shortpath}
Procedure {\tt UniquePath} computes a unique shortest path 
from $v$ to $w$, independent of the agent computing it. 
\end{lemma}

\begin{proof}
All shortest paths between two points inside a terrain can be computed as in \cite{HS97}.
The path computed by the call of {\tt UniquePath}($v$, $w$) is a shortest path, since it is composed, by construction, of parts of shortest paths between $v$ and $w$. The path is computed in a deterministic way without using the compass direction of the agent or the unit of length of the agent. Hence, it is unique.
\end{proof}

\subsection{Coherent compasses}

If agents have a map and coherent compasses, then they can easily agree on one of their two starting positions and meet at this point at cost 
$D$, which is optimal. This is done by the following Algorithm RVCM (rendezvous with a map and coherent compasses).

\fbox{
   \begin{minipage}{0.98\textwidth}
   {\bf Algorithm} $\RVMN$

Let $v$ be the northernmost of the two starting positions of the agents. If both agents have the same latitude, let $v$ be the easternmost of them. Let $w$ be the other starting position. The agent starting at $v$ remains inert. The agent starting at $w$ computes the path 
$p=\mbox{{\tt UniquePath}}(w,v)$ and moves along $p$ until $v$.

\end{minipage}
}

\begin{theorem}\label{th:rvmn}
Algorithm $\RVMN$ guarantees rendezvous at cost $D$,
for any two agents with a map and coherent compasses, in any terrain.
\end{theorem}

\begin{proof}
The position $v$ computed by the two agents is the same, since they have coherent compasses.
The agents will eventually meet in $v$. The cost of rendezvous is $D$, since $p$ is of length $D$.
\end{proof}

\subsection{Incoherent compasses}

\subsubsection{Terrains without obstacles}

In an empty polygon there is a unique shortest path between starting positions of the agents~\cite{AMP91}, and agents with a map can meet in the middle of this path at cost $D$, which is optimal. This is done by Algorithm RVM (rendezvous with a map, without obstacles).

\fbox{
   \begin{minipage}{0.98\textwidth}
\noindent{\bf Algorithm} $\RVM$\\
The agent computes the (unique) shortest path between the starting positions of the two agents. Then, it moves along this shortest path
until the middle of it.
\end{minipage}
}

\begin{theorem}
Algorithm $\RVM$ guarantees rendezvous at cost $D$
for any two agents with a map, in any terrain without obstacles.
\end{theorem}

\begin{proof}
In a polygon without obstacles, the shortest path between two points is unique and can be 
computed as in~\cite{HS97}.
The two agents will eventually meet in the middle of this shortest path. The cost of rendezvous is 
$D$, since the path is of length $D$. 
\end{proof}

\subsubsection{Terrains with obstacles}

This is the first of the two scenarios where agents cannot always predetermine a meeting point. Therefore they compute a common embedding of a ring on which they are initially situated, and
then each agent executes the rendezvous algorithm from~\cite{DGKKPV} for this ring. Rendezvous is guaranteed to occur on the ring, but the meeting point depends on the walks of the agents determined by the adversary. This is done by Algorithm RVMO (rendezvous with a map, with obstacles).

\fbox{
   \begin{minipage}{0.98\textwidth}
\noindent{\bf Algorithm} $\RVMO$\\

\textbf{Phase 1: computation of the embedding\footnotemark[1] $\cR$ of a ring of size $4$.}
\footnotetext[1]
{This embedding is not necessarily homeomorphic with a circle, it may be degenerated.}

Let $v$ be the starting position of the agent and let $w$ be the starting position of the other agent.
The agent computes the embedding $\cR$ of a ring, composed of four nodes $v$, $a$, $w$ and $b$,
where $a$ is the midpoint of  {\tt UniquePath}($v$, $w$), $b$ is the midpoint of {\tt UniquePath}($w$, $v$),
and the four edges are  the respective halves of these paths.

\textbf{Phase 2: rendezvous on $\cR$.}

This phase consists in applying the rendezvous algorithm from~\cite{DGKKPV} for ring $\cR$, 
whose size (four) is known to the agents.

\end{minipage}
}

\begin{theorem}\label{th:rvmo}
Algorithm $\RVMO$ guarantees rendezvous at cost $O(D| l|)$
for arbitrary two agents with a map, in any terrain.
\end{theorem}

\begin{proof}
Let $a_1$ and $a_2$ be the two agents that have to meet. The embedding $\cR$ of the ring is the same for the two agents by Lemma~\ref{lem:shortpath}. The algorithm from~\cite{DGKKPV} guarantees rendezvous and has complexity expressed in terms of the total number of edge traversals by the agents before rendezvous occurs, equal to $O(n|l|)$, where $n$ is the number of nodes of the ring and $l$ is the smaller of the two labels of the agents. Since the ring has size four and each of its edges has length $D/2$, the total cost of rendezvous is 
$O(D|l|)$.
\end{proof}

The following lower bound shows that the cost of Algorithm RVMO cannot be improved for some terrains. Indeed, it implies that for all $D>0$, there exists a polygon with a single obstacle, for which the cost of any rendezvous algorithm for two agents, starting at distance $D$, is $\Omega(D|l|)$. 

\begin{theorem}\label{lem:claim}
For any  rendezvous algorithm $A$, for any $D>0$, and for any integers $k_2\geq k_1>0$, there exist two labels $l_1$ and $l_2$ of lengths at most $k_1$ and at most $k_2$, respectively, and a polygon with a single obstacle of perimeter $2D$, such that algorithm $A$ executed by agents with labels $l_1$ and $l_2$ starting at distance $D$, requires cost $\Omega(Dk_1)$. This holds even if the two agents have a map.
\end{theorem}

\begin{proof}
The idea of the proof is based on an argument from \cite{DFKP}.
For $y>0$, we consider a terrain $\cT$ that is a hexagon of side $y+2$ with one hexagonal obstacle of side $y$ with the same center. The two agents start at positions $u$
and $v$ in $\cT$, as depicted in Figure~\ref{fig:cercle}. The compasses of agents point 
in opposite directions. Observe that $D= 3y$. We call \emph{slices} the six trapezoids bounded by two corresponding parallel sides of the two hexagons and by the segments linking the corresponding vertices of the hexagons. To avoid ambiguity, we say that an agent in the segment shared by two slices is in the first of them in clockwise order. Note that agents start in two different slices with two slices in between.

\begin{figure}[h]
	\begin{center}
	\includegraphics{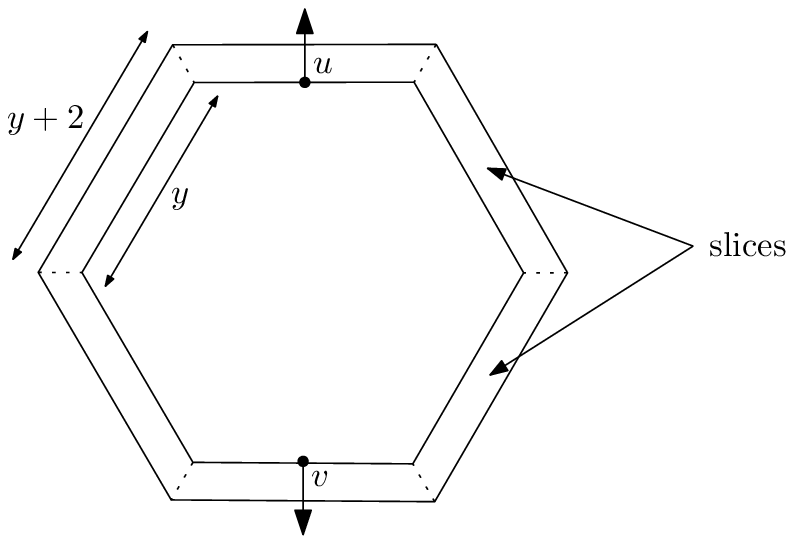}
	\caption{Terrain $\cT$}
	\label{fig:cercle}
	\end{center}
\end{figure}

Fix a rendezvous algorithm $A$. We assume that both agents always move at the same constant speed. We divide the execution of algorithm $A$ into periods during which each agent traverses a distance $y$. During any period, an agent can only visit the slice where it starts the period and one of the two adjacent slices. The behavior of an agent with label $l$, running algorithm $A$, yields the following sequence of integers from the set $\{-1, 0, 1\}$, called the behavior code. The $i$-th term of the behavior code of an agent is $-1$ if the agent ends period $i$ in the slice preceding (in clockwise order) the slice in which it began the period, $1$ if it ends period $i$ in the slice following it (in clockwise order), and $0$ if it begins and ends period $i$ in the same slice. Due to the symmetry of the figure and to opposite compasses an agent with a given label has the same behavior code if it starts at point $u$ or at point $v$. Note that two agents with the same
prefix of length $k_1$ of their behavior codes  cannot accomplish rendezvous during the first $k_1$ periods, since they start separated by at least two slices, and they cannot be in the same slice during any period.

There are less than $3^{k_1/2} < 2^{k_1}$ behavior codes of length at most $k_1/2$. Hence it is possible to pick two distinct labels $l_1$ and $l_2$ of lengths at most $k_1$, respectively, such that the prefix of length $k_1/2$ of their behavior codes is the same. For these labels,  algorithm $A$ does not accomplish rendezvous before both agents have travelled a distance $yk_1/2=\Omega(Dk_1)$. 
\end{proof}

\section{Rendezvous without a map}

\subsection{Coherent compasses}

It turns out that agents can recognize the outer boundary of the terrain even without a map. Hence, if their compasses are coherent, they can identify a uniquely defined point on this boundary and meet in this point. This is done by Algorithm RVC (rendezvous with coherent compasses) at cost $O(P)$. 

\fbox{
   \begin{minipage}{0.98\textwidth}
\noindent{\bf Algorithm} $\RVN$\\

From its starting position $v$, the agent follows the half-line $\alpha$ pointing to the North, as far as possible. When it hits the boundary of a polygon $\cP$ (i.e., either the external boundary of the terrain or the boundary of an obstacle), it traverses the entire boundary of $\cP$. Then, it computes the point $u$ which is the farthest point from $v$ in $\cP\cap \alpha$. It goes around $\cP$ until reaching $u$ again and progresses on $\alpha$, if possible. If this is impossible, the agent recognizes that it went around the boundary of $\cP_0$. It then computes the northernmost points in $\cP_0$. Finally, it traverses the boundary of $\cP_0$ until reaching the easternmost of these points. 
\end{minipage}
}

\begin{theorem}
Algorithm $\RVN$ guarantees rendezvous at cost $O(P)$ for any two agents
with coherent compasses, in any terrain.
\end{theorem}

\begin{proof}
The first phase of the algorithm that consists in reaching $\cP_0$ and making the tour of the boundary of $\cP_0$ costs at most $3P$, since the boundary of each polygon of the terrain is traversed at most twice and the total length of parts of $\alpha$ inside the terrain is at most $P$. Reaching the rendezvous point costs at most $P$. The agents will eventually meet in the easternmost of the northernmost points of $\cP_0$, since they have coherent compasses and this point is unique.
\end{proof}

The following lower bound shows that the cost of Algorithm RVC is asymptotically optimal,  for some polygons even without obstacles.
This lower bound  $\Omega(P)$ holds even if the distance $D$ between starting positions of agents is bounded and if their compasses are coherent.

\begin{theorem}\label{th:doublepie}
There exists a polygon of an arbitrarily large perimeter $P$, for which
the cost of any rendezvous algorithm  for two agents with coherent compasses starting at any distance $D>0$, is $\Omega(P)$.
\end{theorem}

\begin{proof}
Consider the polygon $\cP'$ obtained by attaching to each side of a regular $k$-gon, whose center is at distance $D/8$ from its boundary, a rectangle of length $3D/8$ and of height equal to the side length of the $k$-gon. The polygon $\cP$ is the polygon obtained by gluing two copies of $\cP'$ by the small side of one of the rectangles, as depicted in Fig.~\ref{fig:camembert}. Let $P$ be the perimeter of the polygon $\cP$. We choose $k=\Theta(P/D)$. There are two types of rectangles in $\cP$, two \emph{passing} ones (they share one side) and the $2k-2$ \emph{normal} ones. 

Consider all rotations of the polygon $\cP$ around its center of symmetry by angles $2\pi i/k$,
for $i=0,\ldots,k-1$. We will prove that any deterministic rendezvous algorithm
requires cost $\Omega(P)$ in at least one of the rotated polygons. 
Each agent starts in the center of a different $k$-gon. We say that an agent has
\emph{penetrated} a rectangle if it has moved at distance $D/8$ inside the rectangle.
In order to accomplish rendezvous, at least one agent has to penetrate a passing rectangle.
Each time one agent penetrates a rectangle, the adversary chooses a rotation, so that all previously penetrated rectangles, including the current one, are normal rectangles. This choice is coherent with the knowledge previously acquired by the agents, since normal rectangles are undistinguishable from each other and an agent needs to penetrate a rectangle in order to distinguish its type. Hence, the two agents have to penetrate a total of $k-1$ rectangles before the adversary cannot rotate the figure to prevent the penetration of a passing rectangle. It follows that at least one of the agents has to traverse a total distance of $\Omega(kD)=\Omega(P)$ before meeting.

\begin{figure}[h]
	\begin{center}
	\includegraphics{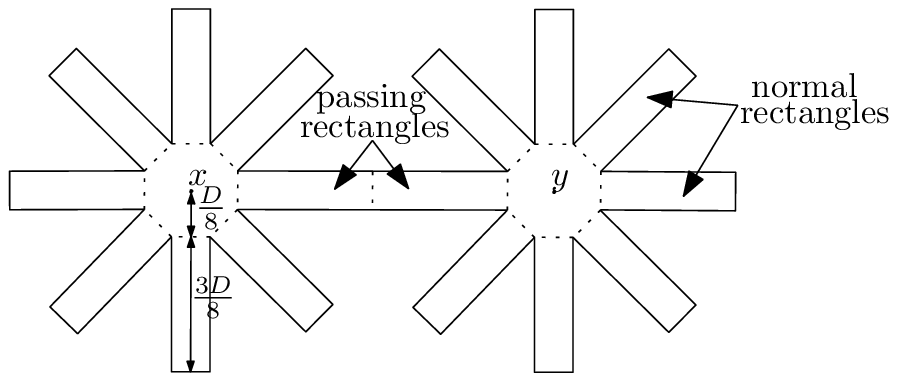}
	\caption{Polygon $\cP$}
	\label{fig:camembert}
	\end{center}
\end{figure}
\end{proof}

\subsection{Incoherent compasses}

\subsubsection{Terrains without obstacles}

In this section, we use the notion of medial axis, proposed by Blum~\cite{Blum67},
to define a unique point of rendezvous inside the terrain. Observe that we cannot use 
the centroid for the rendezvous point since, as we also consider non-convex terrains, the centroid is not necessarily inside the terrain. The \emph{medial axis} $M(\cP)$ of a polygon $\cP$ is defined as the set of points inside $\cP$ which have more than one closest point on the boundary of $\cP$. Actually, $M(\cP)$ is a planar tree contained in $\cP$, in which nodes are linked by either straight-line segment or arcs of parabolas~\cite{Prep77}. We define the \emph{medial point} of a polygon $\cP$ as either the central node of $M(\cP)$ or the middle of the central edge of $M(\cP)$, depending on whether $M(\cP)$ has a central node or a central edge. Remark that the medial point of $\cP$ is unique and is inside $\cP$. The medial axis of a polygon $\cP$ can be computed as in~\cite{CSW95}. Algorithm RV (rendezvous without obstacles,
without a map and with possibly incoherent compasses) determines the unknown (empty) polygon and guarantees meeting in its medial point.

\begin{center}
\fbox{%
   \begin{minipage}{0.98\textwidth}
\noindent{\bf Algorithm} $\RV$\\

At its starting position, the agent chooses an arbitrary half-line $\alpha$ which it follows until it hits the boundary of the polygon $\cP_0$. It traverses the entire boundary of $\cP_0$ and computes the medial point $v$ of $\cP_0$. Then, it moves to $v$ by a shortest path and stops.

\end{minipage}
}
\end{center}

\begin{theorem}\label{th:rv}
Algorithm $\RV$ guarantees rendezvous at cost $O(P)$
for any two agents, in any terrain without obstacles.
\end{theorem}

\begin{proof}
The cost of reaching the boundary of $\cP$ and completing a tour of it is at most $2P$.
The agent can compute the medial point of the polygon and reach it at cost at most $P$. The two agents will eventually meet at the 
medial point, since it is unique.
\end{proof}

The lower bound from Theorem \ref{th:doublepie} shows that the cost of Algorithm RV cannot be improved for some polygons.

\subsubsection{Terrains with obstacles}

Our last rendezvous algorithm, Algorithm $\RVO$, works for the hardest of all scenarios: rendezvous with obstacles, no map, and possibly incoherent compasses. Here again it may be impossible to predetermine a meeting point. Thus agents identify a common cycle and meet on this cycle. The difference between the present setting and that of Algorithm $\RVMO$, where a map was available, is that now  agents may start
outside of the common cycle and have to reach it before attempting rendezvous on it. Also the common cycle is different: rather than being
composed of two shortest paths between initial positions of the agents (a map seems to be needed to find such paths), it is the boundary of a (possible) obstacle $\cO$ in which the medial point of the outer polygon is hidden. These changes have consequences for the cost of the algorithm.
The fact that the medial point of the outer polygon has to be found and  the obstacle  $\cO$  has to be reached is responsible for the summand $P$ in the cost. The  only bound on the perimeter of this obstacle is $x$. Finally,  the fact that the adversary may delay the agent with the smaller label and force the other agent to make its tours of obstacle
 $\cO$  before the agent with the  smaller label even reaches the obstacle, is responsible for the summand $x|L|$, rather than $x|l|$, in the cost.

A \emph{cycle} is a polygonal path whose both extremities are the same point.
A \emph{tour} of a cycle $\cC$ is any sequence of all the segments of
$\cC$ in either clockwise or counterclockwise order starting from a vertex of $\cC$.
By extension, a \emph{partial tour} of $\cC$ is a path which is a subsequence of a tour of $\C$ with 
the first or the last segment of the subsequence possibly replaced by a subsegment of it.

\begin{center}
\fbox{%
   \begin{minipage}{0.98\textwidth}
\noindent{\bf Algorithm} $\RVO$\\

\textbf{Phase 1: Computation of the medial point of $\cP_0$}

At its starting position $z$, the agent chooses an arbitrary half-line $\alpha$ which it follows as far as possible. When it hits the boundary of a polygon $\cP$, it traverses the entire boundary of $\cP$. Then, it computes the point $w$ which is the farthest point from $z$ in $\cP\cap \alpha$. It goes around $\cP$ until reaching $w$ again and progresses on $\alpha$, if possible. If this is impossible, the agent recognizes that it went around the boundary of $\cP_0$. The agent computes the medial  point $v$ of $\cP_0$. 

\textbf{Phase 2: Moving to the medial point of $\cP_0$}

Let $u$ be the current position of the agent. The agent follows the segment $\overline{uv}$ as far as possible. Similarly as in the first phase of the algorithm, if the agent hits a polygon $\cP$, it traverses the entire boundary of $\cP$. Then, it computes the point $w$ which is the farthest point from $u$ in $\cP\cap \overline{uv}$. It goes around $\cP$ until reaching $w$ again and progresses on $\alpha$, if possible. If this is impossible and if the point $v$ has not been reached, the agent recognizes that the point $v$ is inside an obstacle $\cO$, and executes phase 3. If the agent reaches $v$, it does not enter phase 3 of the algorithm and stops.

\textbf{Phase 3: Rendezvous around the medial obstacle of the terrain}

The agent goes around the obstacle $\cO$ until it reaches a vertex $s$. 
The agent produces the modified label $\mu^*$ consisting of the binary representation of the label $\mu$ of the agent followed by a 1 and then followed by $|\mu|$ zeros.
This phase consists of $|\mu^*|$ stages. In stage $i$,
the agent completes two tours of the boundary of $\cO$, starting and ending in $s$, clockwise if the $i$-th bit of $\mu^*$ is 1 and counterclockwise otherwise. 

\end{minipage}
}
\end{center}

Let $\overline{u_1u_2}$ and $\overline{u_2u_3}$ be consecutive segments in clockwise order 
(resp. counterclockwise order) of a cycle. For a given walk $f$ of an agent $a$, we say that the agent \emph{traverses in a clockwise way} (resp. \emph{in a counterclockwise way}) a vertex $u_2$ of a cycle at time $t$ if $f(t)=u_2$ and there exist positive reals $\epsilon_1$ and $\epsilon_2$ and points $y$ and $z$ such that $y=f(t-\epsilon_1)$ is an internal point of $\overline{u_1u_2}$, $z=f(t+\epsilon_2)$ is an internal point of $\overline{u_2u_3}$ and the agent walks in $\overline{u_1u_2}\cup\overline{u_2u_3}$ during the time period $[t-\epsilon_1,t+\epsilon_2]$. 

Before establishing the correctness and cost of Algorithm $\RVO$, we need to show the following two lemmas.

\begin{lemma}\label{lem:traver}
Consider two agents on cycle $\cC$. Suppose that one agent executes a tour of $\cC$ in some sense of rotation, starting and ending in $v$. If during the same period of time, the other agent either traverses $v$ for the first time in the other sense of rotation or does not traverse it at all, then the two agents meet.
\end{lemma}

\begin{proof}
Let $f_1$ and $f_2$ be the walks of agents $a_1$ and $a_2$, respectively.
Let $t'$ be the moment when agent $a_1$ starts its tour of $\cC$ at some vertex $v$. 
Let $t''$ be the moment when agent $a_1$ ends its tour, if agent $a_2$ does not traverse $v$ in the same period of time, or, otherwise, the first moment after $t'$ when agent $a_2$ traverses $v$.
We cut cycle $\cC$ at vertex $v$ obtaining the path $p$ with extremities $v'$ and $v''$ that are copies of $v$. The walks $f_1$ and $f_2$, during the time period $[t',t'']$, can be transposed in $p$, since neither of the two agents traverses $v$ during the period $(t',t'')$. 
For any $t\in[t', t'']$, let $d_i(t)$ be the distance of agent $a_i$ from $v'$ at time $t$, counted on $p$. 
The two functions $d_1$ and $d_2$ are continuous, since the walks of both agents on $p$ are continuous. Notice that, since the first traversal of $v$ by agent $a_2$ may be only in the sense of rotation opposite to that of agent $a_1$, we have $f_1(t')=v'$ and either $f_2(t'')=v'$ or $f_1(t'')=v''$. Let $\delta(t)=d_1(t)-d_2(t)$. We have $\delta(t')=d'\leq 0$ and $\delta(t'')=d''\geq 0$, since $d_1(t')=0$ and $d_1(t'')\geq d_2(t'')$. The function $\delta$ is thus a continuous function from the interval $[t', t'']$ onto some interval $[c',c'']$, where $c'\leq d'$ and $c''\geq d''$. Since $0$ belongs to the interval  $[c',c'']$, there must exist a moment $t$ in the interval  $[t', t'']$, for which $\delta(t)=0$. For this moment, $f_1(t)=f_2(t)$ and the rendezvous occurs.
\end{proof}

\begin{lemma}\label{lem:cycles}
Consider two agents on a cycle $\cC$ and let $k\geq 0$ be an integer. If an agent executes either a partial tour of $\cC$ followed by at most $k$ tours of $\cC$, or at most $k$ tours of $\cC$ followed by a partial tour of $\cC$, while the second agent executes $k+2$ tours of $\cC$, then the two agents meet.
\end{lemma}

\begin{proof}
Assume for contradiction that the two agents never meet. During each tour of $\cC$
by the second agent, the first agent has to traverse the starting position $v$ of the second agent,
in view of Lemma~\ref{lem:traver}.
Hence, the first agent has traversed $k+2$ times vertex $v$. Notice that an agent cannot
 traverse $v$ without executing a tour of $\cC$ as $v$ is an extremity
of a segment of its route. Hence the first agent has completed at least $k+1$ complete tours of $\cC$ starting and ending at $v$. Finally, the first agent has started executing its tours
at point $v$, a contradiction.
\end{proof}

\begin{theorem}
Algorithm $\RVO$ guarantees rendezvous at cost $O(P+x|L|)$
for any two agents in any terrain for which $x$ is the largest perimeter of an obstacle.
\end{theorem}

\begin{proof}
Let $a_1$ and $a_2$ be the two agents that have to meet. The first phase of the algorithm that consists in reaching $\cP_0$ and making the tour of the boundary of $\cP_0$ costs at most $3P$, since the boundary of each polygon of the terrain is traversed at most twice and the total length of parts of $\alpha$ inside the terrain is at most $P$. For the same reason as in phase 1, the total cost of phase 2 is at most $3P$.

If the medial point of $\cP_0$ is inside the terrain, then the agents meet at the end of phase 2 at total cost of at most $12P$. Otherwise, both agents eventually enter phase 3 of the algorithm and they are on the boundary of the obstacle $\cO$ containing the medial point of $\cP_0$. The cost follows from the fact that each agent travels a distance $O(x|L|)$ in phase 3. Indeed, each agent executes at most $2|L|+1$ stages and each stage costs at most $2x$. Hence it remains to show that rendezvous occurs in this case as well.

Assume for contradiction that the two agents never meet. Notice that the modified label $l^*$ cannot be the suffix of the modified label $L^*$. Indeed, if $|l^*|=|L^*|$ then the two labels are different since $l\neq L$, and otherwise the second part of $l^*$, consisting of 1 followed by $| l |$ zeros, cannot be the suffix of $L^*$. Hence, there exists an index $i$ such that the $(|l^*|-i)$-th bit of $l^*$ differs from the $(|L^*|-i)$-th bit of $L^*$. We call \emph{important} stages the $(|l^*|-i)$-th stage of the agent with label $l$ and the $(|L^*|-i)$-th stage of the agent with label $L$.

For $j=1,2$, let $t_j$ be the moment when agent $a_j$ enters its important stage and let $t'$ be the first moment when both agents have finished the execution of the algorithm. Suppose by symmetry
that $t_1\leq t_2$, i.e., agent $a_1$ was the first to enter its important stage. Then $a_2$ must have entered its important stage during the first tour of the important stage of $a_1$. Otherwise, agent $a_2$ would have completed $2i+2$ tours between $t_2$ and $t'$, while agent $a_1$ would have completed at most $2i+1$ tours. Hence, the two agents would have met in view of Lemma~\ref{lem:cycles}. Hence, from the time $t_2$, agent $a_2$ completes one tour in some sense of rotation, starting and ending at a vertex $v$, while agent $a_1$ either traverses $v$ for the first time in the other sense of rotation or does not traverse it at all. Hence by Lemma~\ref{lem:traver}, the two agents meet.
\end{proof}

The following result gives a lower bound matching the cost of Algorithm RVO.

\begin{theorem}\label{lower}
There exist terrains for which the cost of any rendezvous algorithm is $\Omega(P+x|L|)$.
This holds for arbitrarily small $D>0$.
\end{theorem} 

\begin{proof}
Since our lower bound is expressed as a sum, in order to prove it, we  show two examples, one in which the first summand is as small as possible and the bound is equal to the other summand, and vice-versa. The first example uses the polygon from Theorem~\ref{lem:claim}: $P$ must be at least $x$ and in this example we have $P=\Theta(x)$ and the lower bound is $\Omega(x|L|)$. 
Indeed, consider two integers $m_2 \leq m_1$. By Theorem~\ref{lem:claim},
applied for $k_1=k_2=m_1$, and for any rendezvous algorithm $A$, there exists a label $L$ of length $m_1$, such that the sum of lengths of segments of the route produced by the execution of $A$ by an agent $a_1$ with label $L$ is $\Omega(xm_1)$. The adversary chooses as the 
initial position of the second agent $a_2$ any point outside a path $p$ of length $\Theta(xm_1)$, which is a prefix of the route of agent $a_1$. This point can be chosen arbitrarily close to the initial position of the first agent.
The label of agent $a_2$ is of length $m_2$. Suppose that the start of agent $a_2$ is delayed by the adversary and occurs when $p$ is entirely traversed by agent $a_1$. The two agents do not meet during this traversal of $p$ by the first agent and so the cost of rendezvous is $\Omega(xm_1)=\Omega(x|L|)$. The second example is given by the proof of Theorem~\ref{th:doublepie}. Indeed, in this example there are no obstacles and hence $x=x|L|=0$, while the lower bound is $\Omega(P)$.
\end{proof}

\section{Discussion of parameters}

We presented rendezvous algorithms, analyzed their cost and proved matching lower bounds in all considered scenarios. However,  it is important to note that the formulas describing the cost depend on the chosen parameters in each case. All our results have the following form.
For a given scenario we choose some parameters (among $D$, $P$, $x$, $l$, $L$), show an algorithm whose cost in any terrain is
$O(f)$, where $f$ is some simple function of the chosen parameters, and  then prove that for some class of terrains any rendezvous algorithm requires cost $\Omega (f)$, which shows that the complexity of our algorithm cannot be improved in general, for the chosen parameters.

This yields the question which parameters should be chosen. In the case of complexities $D$ and $\Theta(P)$, this choice does not seem controversial, as here $D$ and $P$ are very natural parameters, and the only ones in these simple cases. However, for the two scenarios with incoherent compasses and with the presence of obstacles, there are several other possible parameters, and their choice may raise a doubt. As mentioned in the introduction, in these two scenarios, distinct labels of agents are necessary to break symmetry, since rendezvous is impossible for anonymous agents. Hence any rendezvous algorithm has to use labels $l$ and $L$ as inputs, and thus  the choice of these labels as parameters seems natural. By contrast, the choice of parameter $x$ may seem more controversial. Why do we want to express the cost of a rendezvous algorithm in terms of the largest perimeter of an obstacle? Are there other natural choices of parameter sets? What are their implications?

Let us start by pondering the second question. It is not hard to give examples of other natural choices of parameters
for the two scenarios with incoherent compasses and with the presence of obstacles. For example, in the hardest scenario
(without a map), we could drop parameter $x$ and try to express the cost of the same Algorithm RVO only in terms of $D$, $P$, $l$, and $L$. Since $x \leq P$, we would get $O(P|L|)$ instead of  $O(P+x|L|)$. Incidentally, as in our lower bound example of terrains we have
$x=\Theta(P)$, this new complexity  $O(P|L|)$  is optimal for the same reason as the former one. 

Another possibility would be
adding, instead of dropping a parameter. We could, for example,  add the parameter $P_e$ which is the length of the external
perimeter of the terrain, i.e., the perimeter of polygon $\cP _0$. Then it becomes natural to modify Algorithm  RVO as follows.
The first two phases are the same. In the third phase, the agent goes around obstacle $\cO$ and compares its perimeter to $P_e$.
If the perimeter of $\cO$ is smaller (or equal), then the algorithm proceeds as before, and if it is larger, 
then the agent goes back to the boundary
of  $\cP _0$ and executes Phase 3 on this boundary instead of the boundary of $\cO$. The new algorithm has complexity $O(P+\min(x,P_e)|L|)$. Its complexity is again optimal because in our lower bound example we can choose the parameter $y=\min(x,P_e)$ and enlarge the largest of the two boundaries by lengthy but thin zigzags. Thus we can preserve the lower bound 
 $\Omega(P+\min(x,P_e)|L|)$, even when $x$ and $P_e$ differ significantly.

The reason why we chose parameters $D$, $P$, $l$, $L$, and $x$ instead of just $D$, $P$, $l$ and $L$, is that complexity $O(P+x|L|)$ shows a certain
continuity of the complexity of Algorithm RVO with respect to the sizes of obstacles: when the largest obstacle decreases, this 
complexity approaches $O(P)$ and it becomes $O(P)$ if there are no obstacles. In this case our algorithm coincides with
Algorithm RV. This is not the case with complexity $O(P|L|)$.
On the other hand, this choice coincides with $O(P+\mbox{ min}(x,P_e)|L|)$
in many important cases, for example for convex obstacles (as then we have $x<P_e$).

It is then natural to ask what happens if we add parameter $x$ in the scenario with incoherent compasses and with the presence of obstacles but with the map. Obviously we could still use Algorithm RVO and get complexity $O(P+x|L|)$. 
 However, our lower 
bound argument in this scenario gives in fact only $\Omega (D+\min(x,D)|l|)$. In our example we had $D=\Theta(x)$ but we only get
$\Omega (D+x|l|)$ even if $D$ is much larger than $x$. On the other hand, if $D$ is much smaller than $x$ we can only get the lower bound $\Omega(D|l|)$ because it matches the complexity of $\RVMO$ in this case. Hence it is natural to ask if there exists a rendezvous algorithm with
cost $O(D+\min(x,D)|l|)$ for arbitrary terrains in this scenario. We leave this as an open question.



\end{document}